\documentclass[reprint,aps]{revtex4-1}
\usepackage{amsfonts}
\usepackage{amssymb}
\usepackage{amsmath}
\usepackage{bbm}
\usepackage{bm}
\usepackage{relsize}
\usepackage{graphicx}
\usepackage{xcolor}
\usepackage{float}
\DeclareMathOperator{\sinc}{sinc}

\newcommand{\Eq}[1]{Eq.\,(\ref{#1})}
\newcommand{\Eqs}[2]{Eqs.\,(\ref{#1}) and (\ref{#2})}
\newcommand{\Eqsss}[3]{Eqs.\,(\ref{#1}), (\ref{#2}), and (\ref{#3})}

\newcommand{\Fig}[1]{Fig.\,\ref{#1}}
\newcommand{\Figs}[2]{Figs.\,\ref{#1} and \ref{#2}}
\newcommand{\Sec}[1]{Sec.\,\ref{#1}}
\newcommand{\Secs}[2]{Secs.\,\ref{#1} and \ref{#2}}

\newcommand{\be}{\begin{equation}}
\newcommand{\ee}{\end{equation}}
\newcommand{\bea}{\begin{eqnarray}}
\newcommand{\eea}{\end{eqnarray}}
\newcommand{\wngph}{\omega_{{n\phantom{'}}}^{g\phantom{'}}}
\newcommand{\wng}{\omega_{{n}}^g}
\newcommand{\wngp}{\omega_{{n'}}^{g'}}
\newcommand{\cng}{c_{{n}}^g}

\date{\today}
\begin{document}
\title{
Accidental and symmetry-protected bound states in the continuum in planar photonic-crystal structures,
studied by the resonant-state expansion
}
\author{Sam Neale}
\affiliation{School of Physics and Astronomy, Cardiff University, Cardiff CF24 3AA, United Kingdom}
\author{Egor A. Muljarov}
\affiliation{School of Physics and Astronomy, Cardiff University, Cardiff CF24 3AA, United Kingdom}

\begin{abstract}
	The resonant-state expansion (RSE) provides a precise and computationally cheap tool to find resonant states in complex systems using the optical modes of a simpler system as a basis. We apply the RSE to a photonic crystal slab in order to identify and analyze its bound states in the continuum (BICs).
We show that the RSE is a useful and reliable method for not only finding the BICs but also for differentiating between accidental and symmetry-protected BICs, as well as for understanding their formation from the basis modes and evolution with structural and material parameters of the system.
The high efficiency of the RSE allows us to track the properties of BICs and other high-quality optical modes, covering the full parameter space of the system in a reasonable time frame.

\end{abstract}
\maketitle
\section{Introduction}
\label{Sec:Intro}
Observable phenomena in optical spectra of an electro-magnetic system can be naturally described by using its resonant states (RSs). Originally introduced in open quantum systems~\cite{SiegertPR39}, the RSs of an optical system are the discrete eigen solutions of Maxwell's wave equation satisfying outgoing boundary conditions~\cite{Weinstein}.  The RSs can communicate with the radiation continuum outside the system, and their electro-magnetic fields are travelling solutions which radiate energy away. The lifetime of a RS is given by the quality factor (Q-factor) which is half of the ratio of the real part to the imaginary part of its frequency. In a planar optical system, such as a dielectric slab, the spectrum of RSs can also include guided modes which have infinite Q-factors and are formed through total internal reflection. Guided modes have real frequencies and form evanescent solutions outside the system, and thus are analogous to bound states in quantum systems.

In a planar photonic-crystal (PC) system, there is a mixing of all possible Bragg harmonics, owing to a spatial periodicity, which couples guided modes to the radiation continuum outside the system~\cite{WhittakerPRB99}.  Intuitively, this would lead to the conclusion that guided modes cannot exist in a photonic crystal since all modes have some pathway to the outside, but in reality, modes can remain localized within the system despite their frequencies lying in the continuum. These are known in the literature as bound states in the continuum (BICs), first proposed by von Neumann and Wigner~\cite{vonNeumann1929} in quantum systems and then studied intensively in optical systems~\cite{Inoue1982,Paddon2000,Pacradouni2000,TikhodeevPRB02,MarinicaPRL08,BulgakovPRB08}, where they have been experimentally observed~\cite{Plotnik2011,Hsu2013}. Ideally, BICs have infinite Q-factors, corresponding to $\delta$-like features in optical spectra and infinitely long lifetimes, and therefore have found applications in lasers~\cite{Kodigala2017,Zhen2013}, sensing~\cite{Yanik2011} and filtering~\cite{Foley2014}. In periodic optical systems and in particular PC slabs, BICs have become the subject of growing theoretical interest in recent years~\cite{NdangaliJMP10,LiSR16,BulgakovPRL17,BulgakovPRA17,AzzamPRL18}, including development of various perturbative approaches~\cite{BlanchardPRA14,YangPRL14,YoonSR15,BlanchardPRB16,NiPRB16,Sadrieva2017,YuanOL17,BulgakovPRA18} and other approximate methods~\cite{BykovPRA19,OvcharenkoPRB20}.

There are two main mechanisms by which BICs are disconnected from the radiation continuum: by virtue of symmetry and by the tuning of parameters such that the radiation from all radiating channels is suppressed. These are known as symmetry-protected (S-BICs) and accidental BICS (A-BICs), respectively~\cite{BulgakovPRA14,Hsu2016,GaoSR16}. In a system that exhibits one or more symmetries, the modes of different symmetry classes completely decouple. This causes the two symmetry classes to have different radiative threshold frequencies so that a mode from one symmetry class remains as a guided mode while lying within the continuum of the other symmetry class and in this way forms an S-BIC. Although these are called BICs in the literature, they are akin to the guided modes in a homogeneous system, since they are separated in frequency from the radiation continuum onset.

A-BICS, however, are not dependent on the symmetry of the system but are instead sensitive to its spatial and geometric parameters. Unlike S-BICS, these modes are decoupled from the continuum by destructive interference occurring at the edges of the system which cancel any outgoing, travelling solutions. In PC systems, in particular, it is only necessary for the influence of radiating Bragg channels to be cancelled in this way for an A-BIC to form. This cancellation of electric fields at the edges of the system
is similar to that observed for BICs in an open Sinai billiard \cite{Pilipchuk2017}, a non-periodic system which nevertheless shares some physical properties with a PC slab, see a more detailed discussion at the end of \Sec{Sec:method} below.

To distinguish between A-BICs and S-BICs in practice, a standard approach in the literature is by breaking the symmetry of the system which ensures that all S-BICs disappear from the spectrum of the RSs. In PC systems this is done by simply allowing a non-vanishing in-plane component of the wave number, equivalent to a non-normal incidence of light~\cite{BulgakovPRA14,GaoSR16}. We propose in this paper a different approach, based on the resonant-state expansion (RSE), which allows us not only to distinguish between A-BICs and S-BICs without breaking the symmetry, but also to formulate mathematical criteria for the different types of BICs.

The RSE is a rigorous method for calculating the RSs of photonic systems~\cite{MuljarovEPL10}.   Using as basis a complete set of the RSs of a simpler system, the RSE makes a mapping of Maxwell's equations onto a linear eigenvalues problem, determining the full set of the RSs of the target system. In addition to higher numerical efficiency~\cite{DoostPRA14,LobanovPRA17}, compared to other computational methods, the RSE provides an intuitive physical picture of resonant phenomena, capable of explaining features observed in optical spectra. So far, the RSE has been applied to finite open optical systems of different geometry and dimensionality~\cite{DoostPRA12,DoostPRA13,DoostPRA14,LobanovPRA19}, as well as to homogeneous and inhomogeneous planar waveguides~\cite{ArmitagePRA14,ArmitagePRA18,LobanovPRA17}. It was generalized to magnetic, chiral and bi-anisotropic optical systems~\cite{MuljarovOL18}, enabling its further application to metamaterials. The RSE has also been used in first perturbation order for PC systems to describe sensing of the refractive index by a periodic array of plasmonic nano-antennas~\cite{WeissPRL16}.
Very recently, we have developed a full version of the RSE for planar PC systems~\cite{Neale2020}, using as basis system a homogeneous dielectric slab and treating the PC structure as a periodic modulation on top of the slab. Comparing results with the asymptotically exact scattering-matrix method~\cite{WhittakerPRB99,TikhodeevPRB02} (also known as Fourier-modal method), we have demonstrated in~\cite{Neale2020} a high level of accuracy and efficiency of the RSE for finding the RSs in PC systems.

In this paper, we use these advantages of the RSE,  as well as  the analytical form of the eigenmode expansion, for studying the origin of BICs in planar PC systems and their evolution with structural and material parameters, such as the thickness of the periodic layer and its permittivity contrast. This allows us to reveal the role of different basis states of the homogenous dielectric slab in the formation of the eigenmodes of the PC slab and to demonstrate the importance of the basis guided modes in the formation of BICs.

Interestingly, the role of different basis states of a homogeneous slab in the formation of BICs in PC structures has been very recently studied within a simple coupled-wave model~\cite{BykovPRA19}, reducing the basis to only a few propagating waveguide modes, coupled via diffraction grating, and neglecting any evanescent channels. Compared to this model, the RSE takes into account all possible channels and basis modes within each channel, striving towards the exact solution. Another most recent paper~\cite{OvcharenkoPRB20} studying BICs in PC slabs uses instead the modes of an infinite PC as basis for treating a finite PC slab. The semi-analytic model developed in that paper also neglect any evanescent solutions and in practice presents a reduced version of the rigourous scattering-matrix method~\cite{WhittakerPRB99,TikhodeevPRB02}.

\section{Method}
\label{Sec:method}

In this section, we summarize for the reader's convenience  the formalism of the RSE applied to planar photonic-crystal (PC) structures~\cite{Neale2020}, outlining the most important results. We start from the matrix eigenvalue problem of the RSE in PC systems, providing in \Sec{Sec:RSE} details of separation of the target system into a basis system and a perturbation and introducing the matrix elements of the perturbation for a one-dimensional spatial periodicity. In \Sec{Sec:Basis} we present the main equations for the RSs and cut modes of the basis system, discussing the origin and importance of the branch cuts. We furthermore provide in \Sec{Sec:Perturbed} explicit expressions for the RS fields, both within and outside the PC slab, which is of crucial importance for understanding and analyzing the phenomenon of BICs, which is done in \Secs{Sec:BICs}{Sec:Results}. This is an essential element of the RSE formalism, which is neither included nor discussed in~\cite{Neale2020}. Also, in our previous publication~\cite{Neale2020}, we only mentioned BICs, not providing any examples of A-BICs or a comparative analysis of different types of BICs, which is now done in \Secs{Sec:BICs}{Sec:Results} of this paper.

\subsection{RSE for planar photonic-crystal structures}
\label{Sec:RSE}

The RSE is mapping Maxwell's equations onto a linear matrix eigenvalue problem,  for planar PC structures taking the following form~\cite{Neale2020}:
\be
	\omega\sum_{g'{n}' }\left( \delta_{gg'}\delta_{{n}{n}'}+ V_{{n}{n}'}^{gg'}\right)c_{{n}'}^{g'}=\wng c_{{n}}^g\,,
	\label{RSE}
\ee
where $\delta_{nn'}$ is the Kronecker delta. The eigenfrequency $\omega$ of each RS of the PC system is the  eigenvalue of \Eq{RSE} corresponding to an eigenvector with components $c_{{n}}^g$ which play the role of expansion coefficients of the RS wave function into  the modes of an unperturbed system having the eigen frequencies $\wng$. Indices $g$ and $n$ label, respectively, the Bragg channels and the basis states within each channel. The full set of basis states consists of subsets of modes corresponding to different Bragg channels. Each subset includes, for the same $g$, both the RSs and cut modes of the basis system, see \Secs{Sec:Basis}{Sec:Perturbed} for more details.

Equation~(\ref{RSE}) is valid for any non-dispersive planar PC structure, finite in one direction and infinitely extended in the other two direction, with one- or two-dimensional periodicity. Also, \Eq{RSE} is valid for any polarization of light and for a rather wide choice the basis system. The limitations for the basis system are such that it has to be uniform in the periodic directions of the PC slab, and that the periodic modulation of the permittivity and/or permeability is included within the volume of the basis system. The target PC system thus differs from the basis system by a perturbation which contains the periodic modulation of the permittivity and/or permeability. This perturbation contributes to \Eq{RSE} in a form of the overlap matrix elements $V_{{n}{n}'}^{gg'}$ between normalized basis states. Note that both the target and the basis system can be dielectric and/or magnetic, however with no frequency dispersion. A version of the RSE developed for PC systems with frequency dispersion of the permittivity, such as a PC slab with a periodic array of plasmonic nanoparticles, has been considered in~\cite{WeissPRL16,WeissPRB17}, although with the basis system being also a PC crystal slab with frequency dispersion.

In this paper, we apply the RSE equation~(\ref{RSE}) to a dielectric PC slab with one-dimensional periodicity and transverse-electric (TE) polarization of light (in the $y$-direction). The permittivity of the PC slab is described by
\be
\varepsilon(x,z)=\epsilon(z)+\Delta\varepsilon(x,z)
\label{basis_pert}
\ee
where $\epsilon(z)$ is uniform and $\Delta\varepsilon(x,z)$ is periodic in the $x$-direction with period $d$, i.e. $\Delta\varepsilon(x+d,z)=\Delta\varepsilon(x,z)$. Choosing, without loss of generality, the basis system to be a slab of thickness $2a$ occupying the region $|z|\leqslant a$ and described by the permittivity profile $\epsilon(z)$, which can be homogeneous or inhomogeneous in the $z$-direction, the matrix elements $V_{{n}{n}'}^{gg'}$ in \Eq{RSE} take the following explicit form:
\be
V_{{n}{n}'}^{gg'}= \int_{-a}^a E_{{n}}^g(z)\Delta\epsilon_{g-g'}(z)E_{{n}'}^{g'}(z)dz\,,
\label{Veps}
\ee
where $E_{{n}}^g(z)$ is the electric field of state $n$ of the basis system
for a given Bragg channel
\be
g=\frac{2\pi m}{d}\,,\ \ \ \ m=0,\,\pm 1\,,\pm 2\,,\dots\,,
\label{Bragg}
\ee
and
\be
\Delta\epsilon_g(z) = \frac{1}{d}\int_0^d \Delta\varepsilon(x,z) e^{-igx} dx
\label{epsg}
\ee
is the $g$-th Fourier coefficient of the periodic perturbation.

The split of the full periodic permittivity $\varepsilon(x,z)$ into the uniform $\epsilon(z)$ and periodic  part $\Delta\varepsilon(x,z)$ is arbitrary. However, it is beneficial to choose the periodic perturbation $\Delta\varepsilon(x,z)$ in such a way that its integral over the period $d$ is zero, i.e.
\be
\Delta\epsilon_0(z)=0\,.
\ee
In this case, $V_{{n}{n}'}^{gg}=0$ for all $n$ and $n'$. Important is however that all the diagonal elements  $V_{{n}{n}}^{gg}$ vanish. Then solving \Eq{RSE} to first perturbation order,
\be
\omega\approx\omega_n^g(1+V_{nn}^{gg})^{-1}=\omega_n^g\,,
\ee
shows no effect of the perturbation on the eigenfrequency in that order, so that the periodic modulation contributes only in the second and higher orders, making the matrix problem \Eq{RSE} quickly converging to the exact solution, as it has been demonstrated numerically in~\cite{Neale2020}. Clearly, this approach is ideal for weak periodic modulations of a homogeneous system, but it also remains efficient for stronger perturbations, like those treated in this work, for which the periodic contrast of the permittivity is of the same order as the permittivity contrast in the basis system (compared to the surrounding medium).

Note that \Eq{RSE} is a generalized eigenvalue problem which can however be reduced~\cite{MuljarovEPL10} to the standard eigenvalue problem,
\be
	\sum_{g'{n}' }\left( \frac{\delta_{gg'}\delta_{{n}{n}'}}{\wng }+ \frac{V_{{n}{n}'}^{gg'}}{\sqrt{{\wngph }}\sqrt{\wngp }}\right)b_{{n}'}^{g'}= \frac{1}{\omega} b_{{n}}^g\,,
\label{RSE2}
\ee
by redefining the eigenvector components as $b_{{n}}^g = c_{{n}}^g \sqrt{\wng/\omega}$. Solving \Eq{RSE2} requires only diagonalization of a complex symmetric matrix having the eigenvalues $1/\omega$. While this transformation is unnecessary, it allows one to use a few times more efficient numerical algorithms than those suited for solving generalized eigenvalue problems, such as \Eq{RSE}.

From \Eq{RSE2} follows the orthogonality of the eigenvectors $b_{{n}}^g$ between different perturbed states, which is standard for any symmetric complex matrix. Requiring also their standard normalization (without complex conjugation), $\sum_{gn} (b_{{n}}^g)^2=1$, results in the following normalization of the expansion coefficients:
\be
\sum_{gn} \wng \left(c_{{n}}^g\right)^2 = \omega\,.
\label{norm}
\ee
A similar normalization of the expansion coefficients in finite optical systems  has been considered in~\cite{LobanovPRA18}, showing that this normalization is equivalent to the proper normalization of the perturbed wave functions in real space~\cite{MuljarovEPL10,MuljarovOL18}. We expect (although leaving it without proof as this is not crucial for the results presented in this work) that the same is true also for the RSs in photonic crystal structures, for which the correct normalization was introduced in~\cite{WeissPRL16,WeissPRB17}. We therefore use the normalization \Eq{norm} for the perturbed electric field illustrated in \Sec{Sec:Results} below.

\subsection{Basis system of the RSE: Homogeneous dielectric slab}
\label{Sec:Basis}

As already mentioned above, the RSE requires a basis system to which a periodic perturbation will be added. According to \Eq{basis_pert}, the basis system is described by the permittivity profile $\epsilon(z)$ which is uniform in the $x$-direction (along which the PC slab is periodic). The basis system determines the basis states which are used in the RSE for expansion of the RS fields of the PC slab. The basis states normally consist of the RSs and cut modes, both specific to the form of $\epsilon(z)$. The basis RSs in turn consist of the waveguide (WG) modes which are optical bound states with real eigenfrequencies and Fabry-Perot (FP) modes which are leaky modes with complex eigenfrequencies, both types of modes being the eigen solutions of Maxwell's wave equation with the permittivity $\epsilon(z)$. Cut modes are not eigen solutions in full sense (they satisfy Maxwell's wave equation but not Maxwell's boundary conditions), but they appear as a result of discretization of the branch cuts of the dyadic Green's function (GF) of the basis system  in the complex frequency plane.

In this paper, we use as basis system a homogeneous dielectric slab of permittivity $\varepsilon_s$ and width $2a$, surrounded by vacuum and infinite in the $x$- and $y$-directions. It is described by the permittivity profile
\be
\epsilon(z)=1+(\varepsilon_s-1)\theta(a-|z|)\,,
\label{slab}
\ee
where $\theta(z)$ is the Heaviside function. Focusing on TE polarization and introducing the wave vector $p$ in the $x$-direction, the full electric field of the basis states, labeled by an integer number $n$, is given by ${\bf E}_n(x,z,t)={\bf e}_y E_n(z)e^{i(px-\omega_n t)}$, where ${\bf e}_y$ is the unit vector in the $y$-direction, and
\be
E_n(z)=
\begin{cases}
	A_n e^{ik_n z} &  z>a\\
	B_n(e^{iq_n z}+(-1)^n e^{-iq_n z}) & |z|\leqslant a\\
	(-1)^n A_n e^{-ik_n z} & z< -a
\end{cases}
\label{RSwf}
\ee
is a scalar wave function which consists of standing waves within the slab ($|z|\leqslant a$) and outgoing waves outside the slab ($|z|> a$).
Here, $A_n$ and $B_n$ are the normalization coefficients, linked to each other via Maxwell's boundary condition of the continuity of the electric field,
\be
A_n e^{ik_n a}=B_n(e^{iq_n a}+(-1)^n e^{-iq_n a})\,,
\ee
and the factor $(-1)^n$ accounts for the mode parity, which can be either even or odd, due to the mirror symmetry of the slab in the $z$-direction, see \Eq{slab}. The normal component of the wave numbers in vacuum, $k_n$, and within the slab, $q_n$,  are linked to the RS eigenfrequency $\omega_n$ and the tangent component of the wave number $p$ (which is conserved) via
\be
\omega_n^2=k_n^2+p^2 \ \ \ {\rm and} \ \ \ \varepsilon_s\omega_n^2=q_n^2+p^2\,,
\label{dispersion}
\ee
which are, respectively,  the light dispersion in vacuum and within the slab. Note that we are using throughout this paper the units in which the speed of light in vacuum $c=1$.

\subsubsection{Resonant states}

The eigenfrequencies $\omega_n$ of the basis RSs are generally complex and are found by solving the secular equation,
\be
	(q_n+k_n)e^{-iq_n a}=(-1)^n(q_n-k_n)e^{iq_n a}\,,
	\label{secular}
\ee
which is obtained from Maxwell's boundary conditions of the continuity of the electric field $E_n(z)$ and its derivative. Combining the latter with the outgoing wave boundary conditions, we obtain
\be
E'_n(\pm a)=\pm ik_n E_n(\pm a)\,.
\ee
The last equation for the field on the slab boundaries demonstrates that the general solution of Maxwell's equations (for example, the GF satisfying the same boundary conditions) is analytic in the complex $k$-plane, where $k$ is the normal component of the wave vector in vacuum, which takes the values $k=k_n$ for the RSs. In the complex $\omega$-plane, however, the light dispersion in vacuum introduces branch cuts due to the square root in the light frequency, $\omega=\pm\sqrt{k^2+p^2}$, with the branch points at $\omega=\pm p$, in this way splitting the frequency plane into two Riemann sheets, with the RSs at $\omega=\omega_n$ distributed between the sheets. This creates a choice for which sheet should be taken into account in any expansion, with the states on the other sheet being discarded. Note that the cuts introduce a continuous contribution to the expansion, on top of the discrete contribution of the RSs on the selected Riemann sheet. In other words, both RSs and states on the cuts are required for completeness. The contribution of the cuts and their discretization is considered in more detail in \Sec{Sec:Cuts} below.

The wave functions of the RSs are normalized in such a way that
\be
2\int_{-a}^a \epsilon(z)E^2_n(z) dz -\frac{E_n^2(a)+E_n^2(-a)}{ik_n}=1\,,
\label{normalization}
\ee
which determines the normalization constants in \Eq{RSwf}:
\be
B_n^{-2}=8(-1)^n\left[\varepsilon_s a+\frac{ip^2}{k_n\omega_n^2}\right]\,,
\label{Bn}
\ee
see~\cite{Neale2020} for derivation of \Eqs{normalization}{Bn}.

\subsubsection{Cut modes}
\label{Sec:Cuts}
As already  mentioned, states on the branch cuts in the complex $\omega$-plane contribute to the completeness and thus have to be taken into account in any expansion. For practical purpose, the continuous contribution of the cuts is discretized, replacing each cut with a series of artificial ``cut modes'' positioned on the branch cut and added to the basis along with the discrete RS. As shown in~\cite{Neale2020}, this discretization has the same effect as the truncation of the infinite countable basis of RSs, and the optimal number of the basis cut modes is about the same as the number of the basis RSs. There is also a lot of freedom in choosing the direction of the cuts going from the branch points at $\omega=\pm p$ to infinity. However, the cuts directed vertically down,
\be
\omega=\pm p - i\lambda\,,\ \ \ \ 0<\lambda<\infty
\ee
(with a positive real $\lambda$), turn out to be close to the optimal ones, almost minimizing the cut contribution.

The cut contribution is evaluated from the analytic properties of the dyadic GF of the homogeneous dielectric slab~\Eq{slab}, by extracting the cut density function:
\be
\sigma_\pm(\omega)=\frac{1}{4 \pi} \frac{k}{(k^2-q^2)\cos(2qa)\pm(k^2+q^2)}\,,
\label{sigma}
\ee
where $k=\sqrt{\omega^2-p^2}$ and $q=\sqrt{\varepsilon_s\omega^2-p^2}$, in accordance with \Eq{dispersion}.
Here the sign of $q$ is arbitrary, but the sign of $k$ (changing to the  opposite across the cut) is taken as its value on the left-hand (right-hand) side of the left (right) cut. The other sign in \Eq{sigma}, $\pm$, in turn, refers to the state parity. Note that the cut density \Eq{sigma} is valid for any direction  of the cuts, not only for the vertical direction used in this work.

The continuous cut contribution is discretized by splitting the cuts into finite number of pieces of length $\Delta \omega_n$ in such a way that the value of $\int_{\Delta \omega_n}\sqrt{|\sigma_\pm(\omega)|}d\omega$ is the same for all pieces. Then the cut mode frequencies $\omega_n$ and the normalization constants $B_n$ standing in the wave function \Eq{RSwf}  are defines as
\be
B_n^2=\int_{\Delta \omega_n}\omega \sigma_\pm(\omega) d\omega=  \omega_n \int_{\Delta \omega_n}\sigma_\pm(\omega) d\omega\,,
\ee
where the parity sign can be encoded with an integer state number $n$, exactly in the same way as for the RSs: \mbox{$\pm = (-1)^n$.}
More details on the cut discretization and the use of the cut modes in the RSE can be found in~\cite{Neale2020}.

As already mentioned, the cut modes are not solutions of Maxwell's equations, as the Maxwell boundary condition of the continuity of $E'_n(z)$ on the slab boundaries is not fulfilled for them. Moreover, the field outside the slab, while formally introduced by \Eq{RSwf} also for the cut modes, is physically not defined for them. However, the field of the basis states outside the basis system is not required in the RSE formalism.

\subsection{RSs of a photonic-crystal slab}
\label{Sec:Perturbed}

The homogeneous slab in vacuum considered in \Sec{Sec:Basis} presents the unperturbed system for the RSE.
The RSs and cut modes of the homogeneous slab are introduced in \Sec{Sec:Basis} for a given fixed wave vector $p$ parallel to the slab. To determine, for the same wave vector $p$, the RSs (and cut modes) of a PC slab, the RSE treats the difference between the PC slab and the homogeneous slab as a perturbation that is not necessarily small. Since this perturbation is periodic (in the $x$-direction), it mixes $p$ with   wave vectors $p+g$ of all possible Bragg replicas, where $g$ is defined by \Eq{Bragg}. This implies that in order to obtain the exact result, we need to take all these Bragg channels into account simultaneously in the basis. We denote these Bragg channels with the upper index $g$, which appears in the basis frequencies $\omega^g_n$ and wave functions $E_n^g(z)$ obtained, respectively, from $\omega_n$ and $E_n(z)$ presented in \Sec{Sec:Basis}, by making a replacement $p\to p+g$ in all the equations for the basis RSs and cut modes.  In theory, all Bragg channels should be included, but in practice, the basis is truncated by some maximum frequency $\omega_{\rm max}$ determining a circle in the complex plane,  within which all the unperturbed modes for all possible channels  are taken into account.

Using this extended basis which includes all the RSs and cut modes for all Bragg channels within the cut-off frequency, $|\omega^g_n|<\omega_{\rm max}$, the RSE matrix equation (\ref{RSE}) is solved for a periodic perturbation of interest, $\Delta\varepsilon(x,z)$, in order to find the optical modes of the target system. Since the RSE equation is a linear eigenvalue problem, the number of the output eigenfrequencies $\omega$ of the PC system is exactly the same as the number of the input states $\omega^g_n$ of the truncated basis. Moreover, as approximately half of the basis states are cut modes, about the same number of cut modes are obtained for the target system as a result of solving \Eq{RSE}. Interestingly, these
perturbed cut modes, representing discretized cuts of the PC slab, are positioned in the complex frequency plane along the cuts of the unperturbed system~\cite{Neale2020}. In particular, the real part of the eigenfrequency takes the same values as for the basis cut modes; however, their imaginary parts  are different, which implies a renormalization of the cut density \Eq{sigma} due to the perturbation. This remarkable property of the RSE, that it conserves the positions of the cuts, allows us in particular to distinguish the cut modes from the physical RSs of the target system. This is usually not achievable by other available numerical approaches~\cite{GrasOL19} also dealing with the cuts of PC systems.

The electric field of a perturbed RS (or a perturbed cut mode) is then given by ${\bf E}(x,z,t)={\bf e}_y E(x,z)e^{i(px-\omega t)}$, where
\be
E(x,z)=\sum_{gn} \cng E_n^g(z)e^{igx} \ \ \ {\rm for}\ \ \  |z|\leqslant a
\label{perturbed}
\ee
is the corresponding scalar wave function {\it within} the basis system. We see that the perturbed wave function  $E(x,z)$ is expressed in \Eq{perturbed} as a superposition of the basis modes $E_n^g(z)$ combined with plane waves $e^{igx}$ of the Bragg channels, with the expansion coefficients $\cng $ and the eigenfrequency $\omega$ being a solution of \Eq{RSE}.

Although the basis RSs are defined by \Eq{RSwf} both within and outside the basis slab, the expansion \Eq{perturbed} for a perturbed mode is valid only within the bounds of the basis system, i.e. for $|z|\leqslant a$. To find the electric field of the perturbed mode outside the basis system, we can use the homogeneity of the outside medium, which allows us to find an explicit analytic solution in terms of plane wave. This solution is exactly matching the field \Eq{perturbed} on the surface of the basis system, i.e. at $z=\pm a$. We therefore find the field {\it  outside} the basis system also in terms of the expansion coefficients $\cng $:
\be
E(x,z)=\sum_{g}e^{igx}	e^{ i\varkappa_g (|z|-a)}\sum_n\cng E_n^g(\pm a) \ \ \ {\rm for}\ \ \  |z|\geqslant a\,,
\label{perturbed waves outside}
\ee
where the sign $+$ ($-$) refers to the region $z \geqslant a$  ($z \leqslant -a$), and
\be
\varkappa_g=\sqrt{\omega^2-(p+g)^2}\,.
\label{kappa}
\ee
Equation~(\ref{kappa}) for the normal component $\varkappa_g$ of the light wave vector in vacuum for the $g$-th Bragg channels again introduces a square-root ambiguity. However, the positions of the cuts for the perturbed system are known and are actually the same as for the basis system, as explained above. This unambiguously determines the following choice of the root in \Eq{kappa}:
\be
\begin{array}{lll}
{\rm Im}\,\varkappa_g \leqslant 0 &  \ \ \ {\rm if} \ \ \ & |{\rm Re}\,\omega|>|p+g|\,,\\
{\rm Im}\,\varkappa_g > 0 &\ \ \  {\rm if}\ \ \  & |{\rm Re}\,\omega|<|p+g|\,,
\end{array}
\label{roots}
\ee
which can be obtained by analytic continuation, using the corresponding values of $\varkappa_g$ on the real frequency axis. For a better understanding of the meaning of \Eq{roots}, let us assume for definiteness that ${\rm Re}\,\omega>0$. This assumption does not impose any limitations as the modes with ${\rm Re}\,\omega<0$ are solutions which are the complex conjugate of their mirror images with respect to the imaginary axis in the complex $\omega$-plane --- this is a general property of the RSs of an optical system, related to its time-inversion symmetry~\cite{MuljarovEPL10}.
Then the first case in \Eq{roots} corresponds to a so-called {\it open} Bragg channel, for which $e^{ i\varkappa_g z}$ represents a wave propagating away from the system in the positive $z$-direction. Such a wave has a constant amplitude if the mode eigenfrequency $\omega$ is purely real (in this case this amplitude is zero in reality, see a discussion in \Sec{Sec:BICs} below), or an exponentially growing amplitude, due to ${\rm Im}\,\varkappa_g <0 $, if the mode eigenfrequency is complex, i.e. Im\,$\omega<0$. Note that this  exponential growth is a typical spatial behaviour of the RSs~\cite{SiegertPR39,TikhodeevPRB02,MuljarovPRB16Purcell}.
The second case in \Eq{roots} corresponds to a {\it closed} Bragg channels, for which the field due to $e^{ i\varkappa_g z}$ is exponentially decaying with the distance from the system,  due to ${\rm Im}\,\varkappa_g >0 $, no matter whether the eigenfrequency $\omega$ is real or complex. Note that the case of ${\rm Re}\,\omega=p+g$, not included in \Eq{roots}, corresponds to cut modes.

We can see from the expansions \Eqs{perturbed}{perturbed waves outside} that the coefficients $c_n^g$ act as amplitudes controlling how much each basis mode contributes to a given perturbed state. This makes it possible, in particular, to determine which basis mode the perturbed state ``originates'' from, that is, which basis mode it would evolve from if the position of the mode were traced out as the contrast of the periodic modulation is increased.

\subsection{Bound states in the continuum}
\label{Sec:BICs}

As well as allowing us to be able to construct the perturbed electric field, the eigenvectors $\cng$ of the RSE allow us to easily identify BICs in the system and to be able to distinguish between S-BICs and A-BICs.
As BICs have localized (i.e. bound) wave functions, their eigenfrequencies $\omega$ have to be purely real, otherwise an excitation of the system into such an optical mode would decay in time, which would in turn require an exponential growth of the wave functions. And vice versa, all real-eigenfrequency modes have to have localized wave functions, which means they can only be bound states. A purely harmonic behaviour in vacuum (with a finite constant amplitude) is not possible for an isolated optical mode at real frequency as this would also mean a flow of energy to the outside of the system. Mathematically, this implies, in accordance with \Eq{roots} and the discussion following it in \Sec{Sec:Perturbed}, that
\be
\sum_n \cng E_{{n}}^g(\pm a)=0
\label{BIC cond}
\ee
for every Bragg channel $g$ satisfying the inequality
\be
|p+g|<|\omega|\,,
\label{channels}
\ee
where $\omega$ is the {\it real} frequency of the mode. This is a general condition for any BIC. Note that in addition to BICs, also guided modes with real frequencies $|\omega|<|p|$ can form in the energy spectrum of a PC slab in vacuum (the last inequality is modified for a PC slab with a substrate or two different substrates replacing vacuum on either side of the slab~\cite{TikhodeevPRB02}).

For S-BICs, however, a stronger condition replaces \Eq{BIC cond}:
\be
	c_n^g=0\,.
\label{S-BIC cond}
\ee
This should be fulfilled for all basis states $n$ for the Bragg channels $g$ selected by the inequality \Eq{channels}. In fact, as already mentioned in \Sec{Sec:Intro}, S-BICs are formed due to the decoupling of modes of different symmetry, so while all modes of one symmetry class couple to the radiation continuum outside the system and become leaky, some modes of the other symmetry class do not and remain bound to the system. All the matrix elements $V_{{n}{n}'}^{gg'}$ between different symmetry classes vanish by symmetry, and the RSE equation (\ref{RSE}) yields immediately \Eq{S-BIC cond}. For example, in the case of a PC slab treated in \Sec{Sec:Results} below, for $p=0$ and mirror symmetry in the $z$-direction, these two symmetry classes are, respectively, even and odd solutions in the periodic $x$-direction. The corresponding even and odd-parity basis modes do not couple to each other, leading to \Eq{S-BIC cond} for the even-mode contribution to the odd-parity states. From this follows, in particular, that \Eq{S-BIC cond} results in S-BICs existing only below the first Bragg channel threshold, i.e. for $|\omega|<2\pi/d$, as it has been demonstrated numerically in \cite{Neale2020}.  One could even argue that S-BICs are simply guided modes of the odd-symmetry class.

Equation (\ref{S-BIC cond}) is a defining characteristics for S-BICs that is not seen in A-BICs which would be otherwise difficult to differentiate. The RSE method reveals how this decoupling occurs mathematically.
A-BICs, on the other hand, can form at any frequency given that the parameters of the system are properly tuned. Unlike S-BICs the values of $\cng$ are not necessarily zero for leaky modes. Instead, according to \Eq{BIC cond}, it is the summation of the basis electric fields at the edges of the system that becomes zero, representing destructive interference. For an A-BIC to form, this destructive interference only needs to occur in channels that contain leaky modes,
i.e. for all $g$ satisfying \Eq{channels}.

The condition for A-BICs \Eq{BIC cond} with generally non-vanishing amplitudes $\cng$ can be also seen as orthogonality of vectors with component $\cng$ and $E_{{n}}^g(\pm a)$, labeled by $n$, for each relevant $g$. If only one value of $g$ contributes (for example $g=0$, for sufficiently small $|\omega|$), and the PC slab possesses a mirror symmetry in the $z$-direction, the same as for the basis system, then the two equations given by \Eq{BIC cond}, with $+a$ and $-a$ in the argument of the basis functions, produce only one (for each parity in the $z$-direction) ``vector orthogonality'' condition for an A-BIC to occur, which is easy to satisfy by tuning a single parameter of the system, as demonstrated in \Sec{Sec:Results} below. However, for a PC slab without mirror symmetry, and for larger frequencies, \Eq{BIC cond} contains two or more conditions of orthogonality of vectors, which are harder to meet and which may require a simultaneous tuning of, respectively, two or more parameters of the system.

As already mentioned in the introduction, the condition for A-BICs \Eq{BIC cond}, provided by the RSE, which physically reflects the phenomenon of destructive interference and cancellation of the field at the edges of the system, has some similarity with a BIC condition developed in the theory of an open
Sinai billiard~\cite{Pilipchuk2017}. The latter presents an interesting example of an open optical system
treated in a rigorous way without introducing the RSs explicitly, but rather mapping Maxwell's wave equation onto a non-Hermitian matrix eigenvalue problem, using as basis the eigenstates of a closed system supplemented with guided and evanescent modes of leaky channels. While the open Sinai billiard is not a periodic system, it has a remarkable similarity with a PC slab, in terms of the existence of open (i.e. leaky) and closed (i.e. evanescent) channels which are analogous, respectively, to the discussed above open and closed Bragg channels of a PC slab. It has been shown in~\cite{Pilipchuk2017} that, neglecting the contribution of evanescent channels, the condition determining BICs in such a system can be formulated in term of a vanishing coupling matrix element between a relevant mode of the closed resonator and the leaky channel. While for some BICs this approximation works very well, in some other cases the contribution of evanescent mode can more significant, as has been also demonstrated in~\cite{Pilipchuk2017}. In PC systems instead, the A-BIC condition \Eq{BIC cond} is exact.

\section{Results}
\label{Sec:Results}

One of the advantages of the RSE method is the speed at which it can calculate the RSs of an optical system within a wide spectral range. This becomes crucial if one needs to change one or several parameters of the optical system in order to optimize its optical properties or to achieve a desired effect. In fact, if the basis system remains the same while varying parameters of the target system, the perturbation matrix can be pre-calculated as it is using the same basis functions, or at least its calculation can be optimized. Then the only computationally expensive element of the RSE is matrix diagonalization. For example, for the perturbations treated in this work with the relative error in the RS frequency of about $10^{-5}$ or lower, one needs for determining the full spectrum of the RSs in a wide spectral range to diagonalize a $4500\times 4500$ matrix which requires only 680 seconds on a standard computer. This allows the RSE to be run several thousand times within a manageable time frame, in order to explore the parameter space of the PC system and to trace the evolution of its optical modes while its structural and/or material parameters are changing. This is particularly important for studying A-BICs, as discussed in \Sec{Sec:BICs}.

In this section, we introduce a planar PC system in a form of a harmonic perturbation on top of a homogenous dielectric slab and vary the parameters of this perturbation (namely, the perturbation strength and width), in order to study symmetry protected and accidental BICs in such a system.

\subsection{Perturbation matrix and parameters of the PC slab}
\label{Sec:Parameters}

The PC system considered in this paper is essentially the same as in~\cite{Neale2020}. It is described by the total permittivity $\varepsilon(x,z)$, given by \Eq{basis_pert}, in which $\epsilon(z)$ is the permittivity of a homogeneous slab, given by \Eq{slab}, and $\Delta\varepsilon(x,z)$ is a perturbation having the following form
\be
\Delta\varepsilon(x,z)=\beta\cos\left({\frac{2\pi}{d}x}\right)\theta(b-|z|)\,,
	\label{perturbation}
\ee
in which $\beta$ is the perturbation strength, i.e. the contrast of the periodic modulation of the permittivity with period $d$, and $b$ is the perturbation widths.

The matrix elements of the perturbation \Eq{perturbation} are calculated according to \Eqsss{Veps}{Bragg}{epsg}, and take the following simple analytic form:
\be
V_{nn'}^{gg'}=B^g_{n}B^{g'}_{n'}\left(\delta_{g,g_1}+\delta_{g,g_{-1}}\right)\beta b Z^{gg'}_{nn'}\,,
\label{Vnm}
\ee
where $g_{\pm1}=\pm 2\pi/d$\,,
\bea	
\begin{aligned}		
Z^{gg'}_{nn'}=&
\left(1+(-1)^{n+n'}\right)\sinc\left[ \left(q^g_{n}+q^{g'}_{n'}\right)b\right]\\
&+\left((-1)^{n}+(-1)^{n'}\right)\sinc\left[ \left(q^g_{n}-q^{g'}_{n'}\right)b\right]\,,
\end{aligned}
\eea
$\sinc z=\sin z/z$, and $q_n^g=\sqrt{\varepsilon_s \omega_n^2-(p+g)^2}$. Here, $\omega_n^g$ and $B_n^g$ are, respectively, the eigenfrequency and the normalization coefficient of a basis RS or a cut mode with the in-plane wave vector $p+g$, in a homogeneous dielectric slab in vacuum, having permittivity $\varepsilon_s$ and  width $2a$.

For the rest of the paper, we fix the following parameters of the system: $p=0$, $\varepsilon_s=6$, and $d=2\pi/5$, and vary the perturbation strength $\beta$ in \Sec{Sec:Evolution} (while keeping fixed the perturbation width at $b=a/2$) and both $\beta$ and $b$ in \Sec{Sec:Other}. A sketch of the system is provided in the inset of \Fig{snakeplot}.

\subsection{Evolution of modes: Symmetry-protected and accidental BICs}
\label{Sec:Evolution}

The evolution of a large number of the RSs with change of the permittivity contrast $\beta$ has been already considered for this system in~\cite{Neale2020}. In this section, we focus only on a pair of the RSs originating from the fundamental guided modes of the homogeneous slab, corresponding to the first-order Bragg channels, i.e. with $g=g_{\pm1}$. Note that as we consider the RSs of the PC slab with the in-plane wave number $p=0$, the Bragg channels with opposite signs of $g$ in the basis are degenerate by symmetry. This degeneracy also manifests itself in the orthogonality of even and odd states along the $x$-axis. As a consequence of this orthogonality, the matrix elements of the perturbation between even and odd basis states vanish by symmetry, thus allowing formation of S-BICs
as discussed in \Sec{Sec:BICs}. Clearly, these S-BICs exist only for zero in-plane vector $p$ that guarantees the mirror symmetry of the electro-magnetic field in the $x$-direction. With any deviation from $p=0$ condition, breaking the symmetry, these S-BICs transform into modes with finite Q-factors.

\begin{figure}
\vskip-5mm
	\includegraphics*[clip,width=0.56\textwidth]{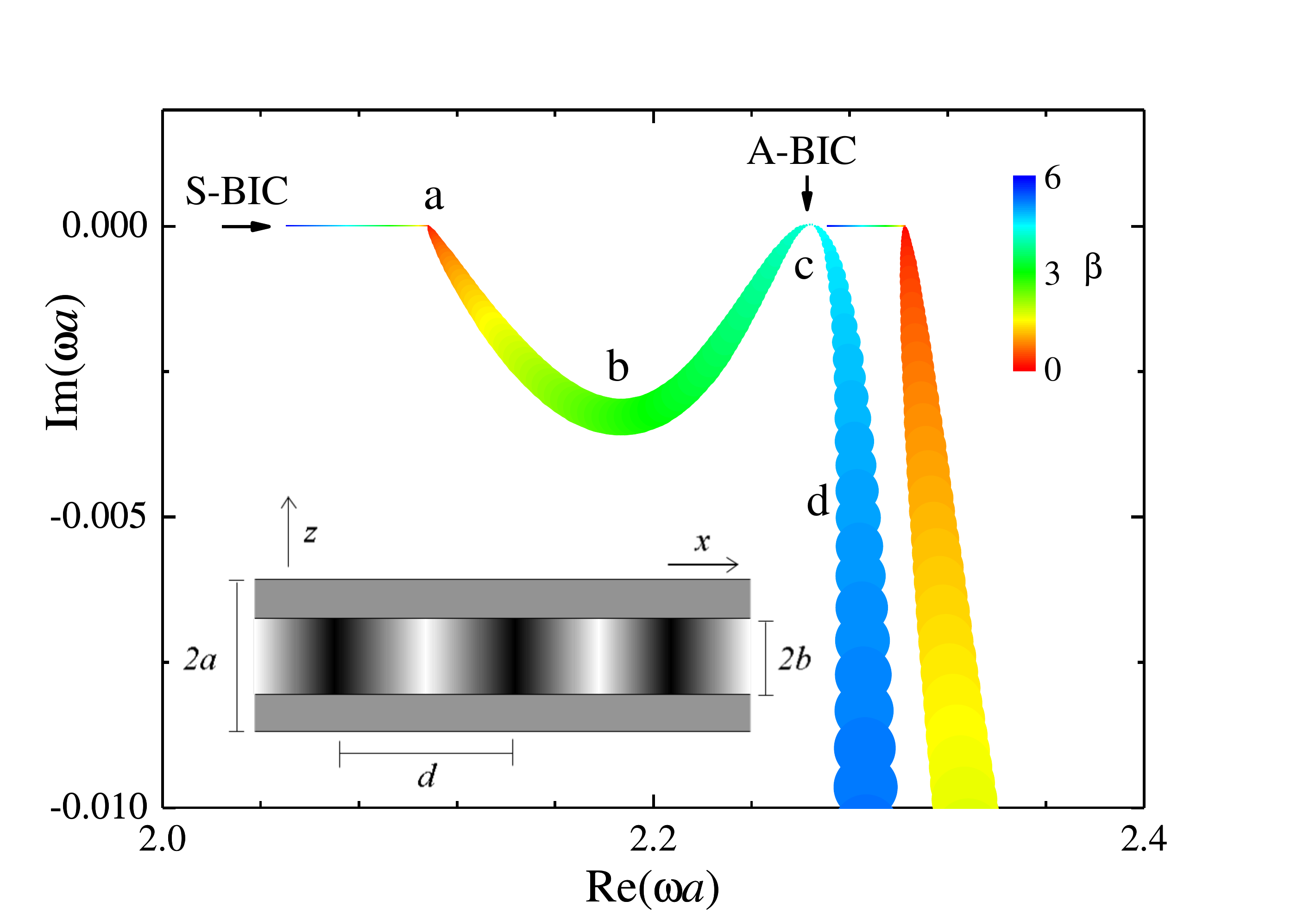}
	\vskip-0mm
	\caption{Evolution of the RS eigenfrequencies (centers of the colored circles) and the zeroth Bragg channel contribution $|C_0|$, defined by \Eq{C0} (circle area), for the amplitude of the periodic modulation $\beta$ (circle color) changing between $\beta=0$ and $\beta=6$. As $\beta$ grows, a doubly degenerate fundamental guided mode of the first Bragg channel (a) splits into a symmetry protected BIC with an infinite Q-factor and a QGM with a finite Q-factor (b,d) which becomes an accidental BIC (c) at $\beta \approx4.34$.
The inset shows a schematic of the target photonic-crystal system. The basis system is a dielectric slab in vacuum, having width $2a$ and permittivity $\varepsilon_s=6$.
}
	\label{snakeplot}
\end{figure}
Figure~\ref{snakeplot} demonstrates the evolution of the fundamental pair of RSs for $p=0$ as the periodic modulation contrast $\beta$ increases.  Being doubly degenerate by symmetry at $\beta=0$ (no periodic modulation), the fundamental guided mode splits for a nonzero $\beta$ into an S-BIC and a quasi-guided mode (QGM), the latter having a rather high but still finite Q-factor.  These two modes are further separated as $\beta$ increases. However, as clear from \Fig{snakeplot}, for a certain value of $\beta$ (at around $\beta\approx4.34$), the QGM transforms into an A-BIC with an infinite Q-factor (corresponding to zero imaginary part of its eigenfrequency).

\begin{figure}
	\includegraphics*[clip,width=0.5\textwidth]{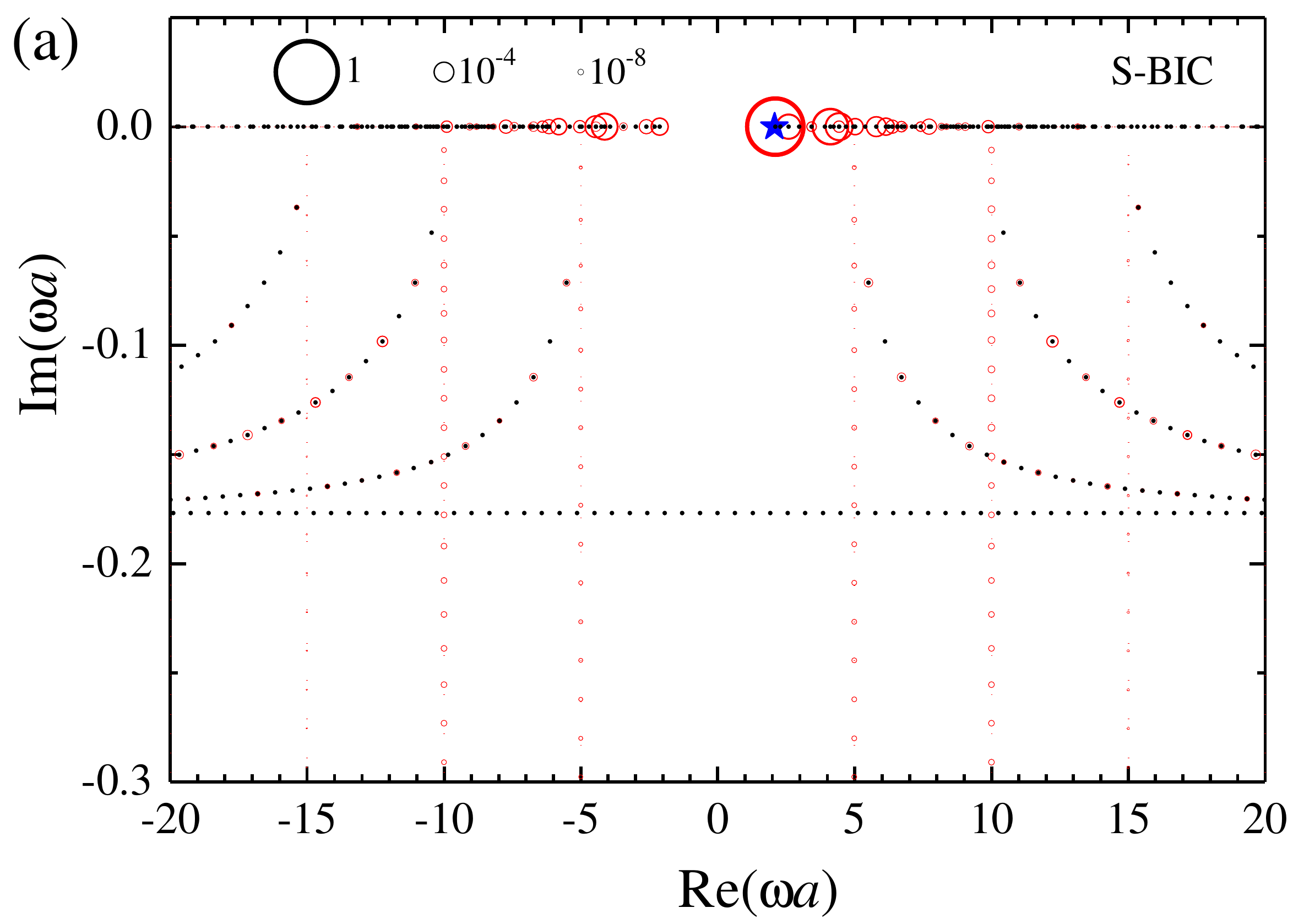}
	\includegraphics*[clip,width=0.5\textwidth]{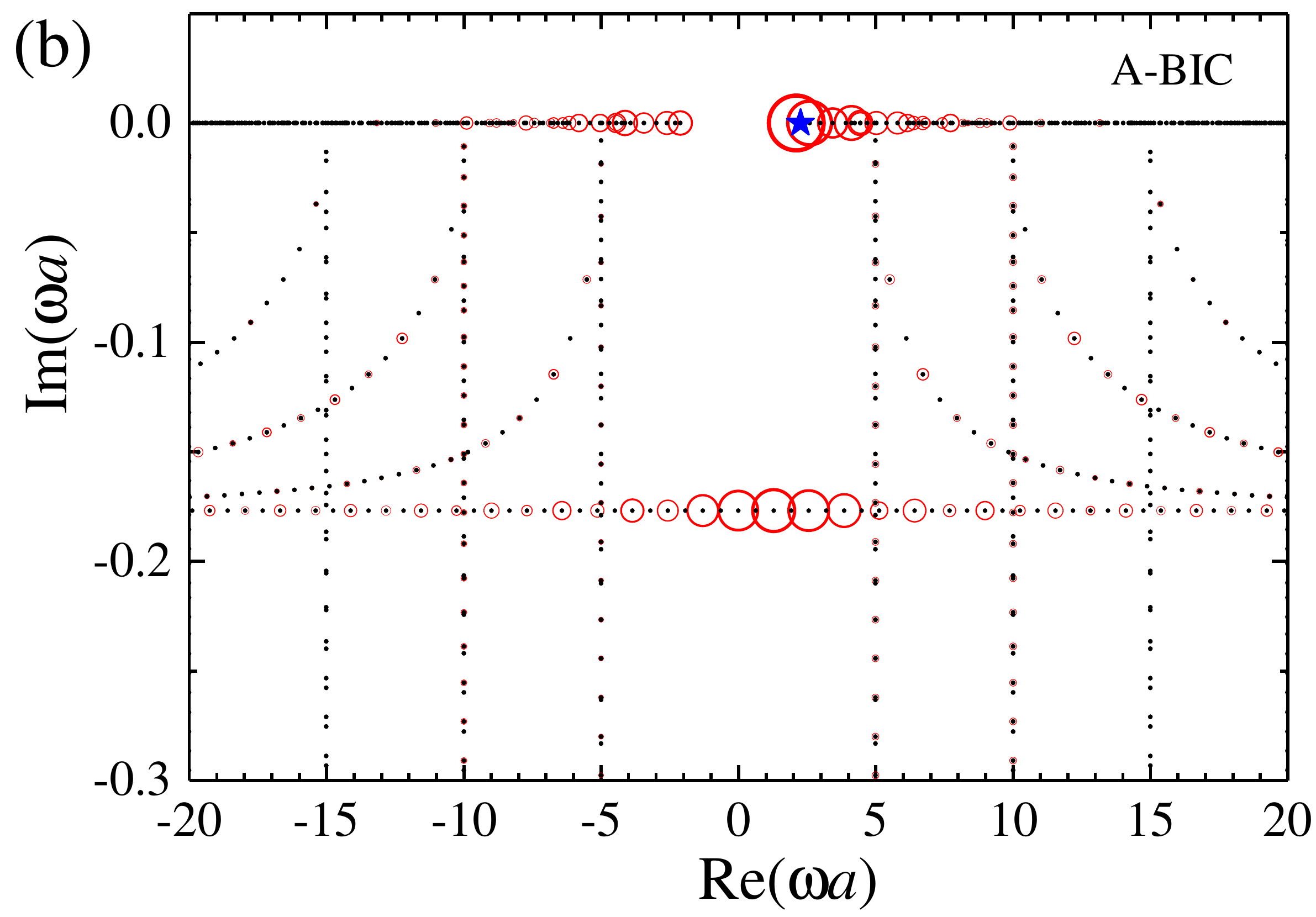}
	\caption{Basis mode contributions (red circles, centered at the basis mode frequencies) for (a) symmetry protected and (b) accidental BIC at $\beta=4.34$, both originating from the same doubly degenerate fundamental guided mode of the homogeneous slab with $m=\pm1$. The circle area is proportional to $\sqrt{|c_n^g|}$ of the basis mode amplitude $c_n^g$, and a key showing the relationship between the circle area and $|c_n^g|^2$ is given by back circles and the numbers next to them. Blue star gives the position of the BIC eigenfrequency and black dots of the basis RSs.}
	\label{contributions}
\end{figure}

The symmetry-protected and accidental BICs, originating from the same fundamental guided mode of the homogeneous slab, are compared in \Fig{contributions}, where the contribution of a large number of basis modes to both states is shown by red circles centered at the basis mode frequencies and having the area of the circle proportional to the modulus of the square root of the expansion coefficient, $\sqrt{|c_n^g|}$. Among several thousand of modes used in the basis, only a limited number of basis states give an appreciable contribution to the perturbed RSs, with a clearly dominating role of the fundamental guided mode, see the largest circle close to the black star, which in turn shows the perturbed mode position on the real axis.

A remarkable difference between the two BICs is a vanishing (non-vanishing) contribution to S-BIC (A-BIC) of the individual basis modes of the zeroth Bragg channel. These basis modes all have the same imaginary part and equidistant separation in the complex frequency plane, see \Fig{contributions}. In accordance with a discussion in \Sec{Sec:BICs} above, all $c_n^0=0$ for the S-BIC, while $c_n^0\neq0$ for the A-BIC, for all modes $n$ of the same (even) parity along $z$-axis. However, the vector with components $c_n^0$ is orthogonal to the vector $E_n^0(a)$ for the A-BIC, as also discussed in \Sec{Sec:BICs}.
The absolute value of the dot product of these two vectors,
\be
C_0=\sum_n c_n^0E_n^0(a),
\label{C0}
\ee
is shown by the circle area in \Fig{snakeplot}, while the center of the circle gives the position of the perturbed RS frequency. Clearly, $C_0$ is vanishing at the A-BIC and is strictly zero for the S-BIC at all values of the periodic modulation amplitude $\beta$.

\begin{figure}
	\includegraphics*[clip,width=0.5\textwidth]{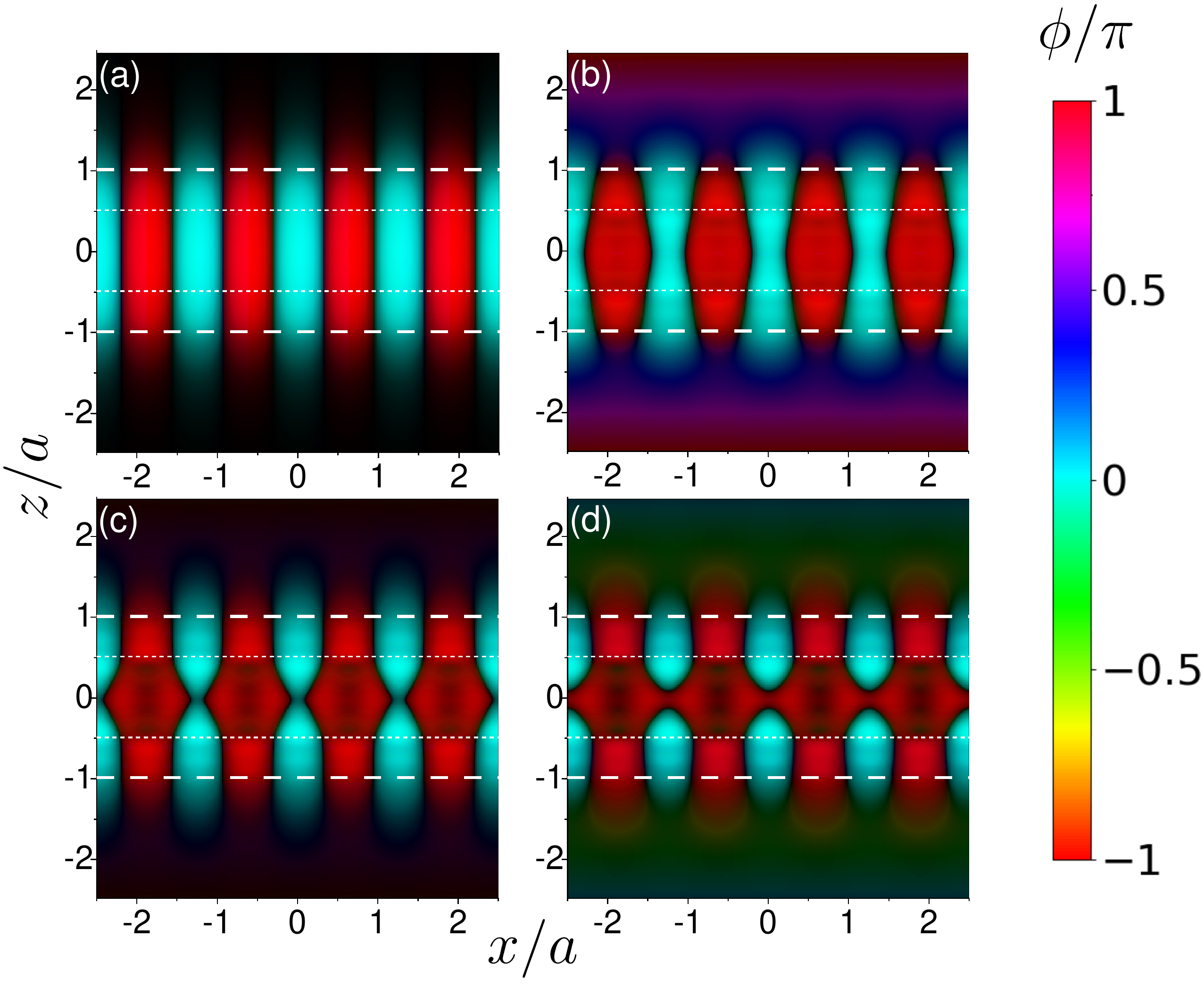}
	\caption{The amplitude (brightness) and phase (color) of the electric field of the fundamental optical mode in \Fig{snakeplot} at the positions labelled a-d, corresponding to the following mode type and permittivity contrast $\beta$: (a)  fundamental basis guided mode of the first Bragg channel at $\beta=0$; (b)  QGM before the BIC, at $\beta=3$; (c)  A-BIC at $\beta\approx 4.34$; (d) QGM after the BIC, at $\beta=5$. The thick (thin) dashed lines indicate the edges of the slab at $z=\pm a$  (periodic perturbation at $z=\pm b$).
}	
\label{e-fields}
\end{figure}

\begin{figure}
	\includegraphics[width=0.5\textwidth]{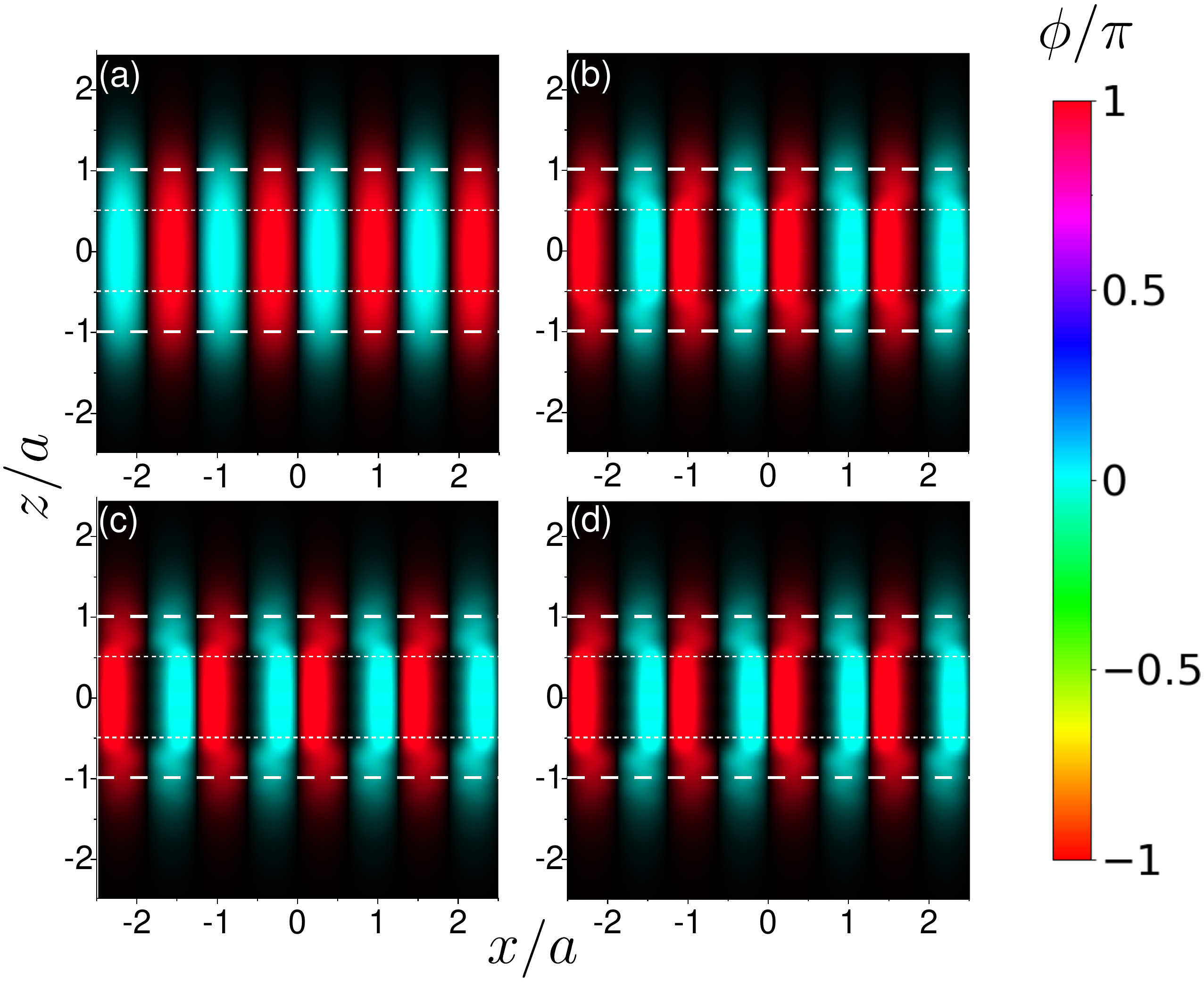}
	\caption{As \Fig{e-fields} but for the S-BIC shown in \Fig{snakeplot}. The values of the permittivity contrast $\beta$ and all other parameters are the same as in \Fig{e-fields}.}
	\label{e-fields S-BIC}
\end{figure}

The transformation of the QGM, passing through the A-BIC as $\beta$ increases, is accompanied by a morphological change of the wave function, which is demonstrated in \Fig{e-fields}.
It shows both the amplitude (intensity) and the phase (color) of the wave function of the QGM  at four different values of $\beta$, starting from $\beta=0$ in \Fig{e-fields}(a) at which the RS coincides with the fundamental guided mode of the homogeneous basis slab. The other three panels (b)--(d) in \Fig{e-fields} show the wave function of the mode before, at, and after the A-BIC.
These four positions are also labeled (with the same letters a-d) in the complex-frequency plot, \Fig{snakeplot}. As the QGM originates from the fundamental guided mode of the homogeneous slab, corresponding to the first Bragg channel, the amplitude of the electric field of this fundamental QGM before it transforms into the A-BIC has only one maximum in the $z$-direction and two maxima per period $d$ in the $x$-direction, the same as for the basis mode. After the A-BIC, however, the field amplitude shows three maxima in the $z$-direction representing a growing contribution from a higher-order basis mode, specifically the third guided mode in the first Bragg channel. Note that the second guided mode does not contribute since it is of the opposite parity to the fundamental mode from which the QGM originates. In the $x$-direction, the  morphology of QGM does not change much: There are always two maxima per period indicating that there is no major change in the contribution from higher-order Bragg channels within this $\beta$ range.

\begin{figure}
	\includegraphics*[clip,width=0.5\textwidth]{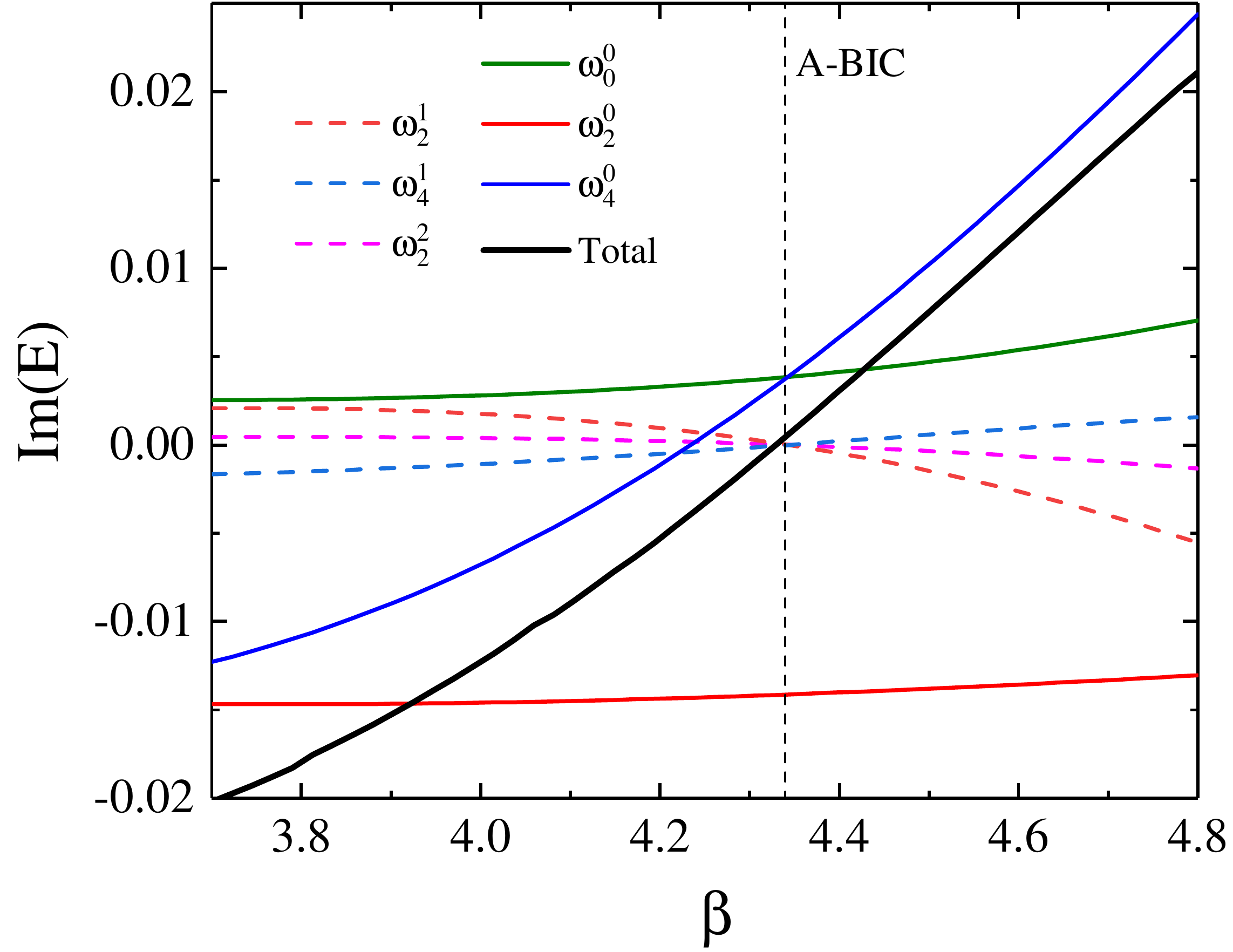}
	\caption{Contributions of basis modes to the imaginary part of the normalized electric field of the fundamental QGM as functions of the permittivity contrast $\beta$. Only the top 6 contributions are shown, corresponding to the basis mode frequencies $\omega_{0}^0=- 0.177i$, $\omega_{2}^0=1.283 - 0.177i$, $\omega_{4}^0=2.565 - 0.177i$, $\omega_{2}^1=2.108$, $\omega_{4}^1=2.605$, and $\omega_{2}^2=4.123$.
The total electric field of the QGM is given by a black thick line.
The vertical dashed line shows the value of $\beta=4.34$ where the QGM becomes the accidental BIC.
}
	\label{e-field contributions}
\end{figure}

It can also be seen that the field quickly decays outside of the system, in a similar way for both the basis mode and the A-BIC, compare \Fig{e-fields}(a) and (c). In fact, since both modes are bound, there are no travelling or exponentially growing solutions outside the system; instead, there are only evanescent waves, unlike the other two cases which are shown in \Fig{e-fields}(b) and (d). For them, a non-vanishing growing field, though very small, is seen in the region outside the system. One can see from the color of the plots that the fields in \Fig{e-fields}(b) and (d) are almost real, which is consistent with the fact that the Q-factor of the QGM shown is very high, see \Fig{snakeplot}. Interestingly, the phase of the wave function is either close to 0, or to $\pi$, with the amplitude of the wave function almost vanishing on lines separating these two phase regions. The  wave functions of the basis guided mode and the A-BIC shown, respectively, in \Fig{e-fields}(a) and (c) are purely real, as expected for any bound states, having also purely real eigenfrequencies. These real wave functions just change their sign along the lines separating the above mentioned 0- and $\pi$-phase areas.

For the same values of $\beta$ as in \Fig{e-fields}, we show in \Fig{e-fields S-BIC} also the evolution of the S-BIC which originates from the same doubly degenerate pair of guided modes as the QGM mode/A-BIC in \Fig{e-fields}. Clearly, the modes shown in \Figs{e-fields}{e-fields S-BIC} are, respectively, of the even and odd parity in the $x$-direction. The fact that the S-BIC does not gain any leakage for any value of $\beta$ is consistent with the dark areas outside the slab in all four panels of \Fig{e-fields S-BIC}. As $\beta$ increases, the S-BIC does not change much within the slab either. In fact, there is only a slight morphological change in the $x$-direction, but no change in the $z$-direction is observed. This suggests that this S-BIC is not communicating much with higher order modes. The overall lack of evolution of the S-BIC is also consistent with the fact that this mode is not moving much in the complex frequency plane as compared to its even counterpart, as it is clear from \Fig{snakeplot}. The S-BIC is also quite isolated in frequency from other basis states which effectively reduces their impact. In fact, as we know, the S-BIC is not communicating at all  with the nearby $g=0$ modes due to symmetry, compare Figs.\,\ref{contributions}\,(a) and \ref{contributions}\,(b).

We now want to see how the summation of complex basis electric fields shown in \Eq{perturbed} create an entirely real field at the A-BIC. To do this,  we plot in \Fig{e-field contributions} the imaginary part of the weighted basis fields $\cng E_{n}^g$ against $\beta$, the sum of which will clearly be zero at the A-BIC.  In order to produce a readable plot we can limit ourselves to just the top few contributors. In this case, only the top six modes with the highest contributions to the A-BIC, well seen in \Fig{contributions}(b), are used.

It can be seen in Fig. \ref{e-field contributions} that the imaginary part of the weighted basis fields from the guided modes all come to zero at the A-BIC. This is expected behaviour since the guided-mode fields and the A-BIC field are purely real, so naturally the corresponding eigenvector components $\cng$ are also real. There are, however,  leaky modes from the $g=0$ Bragg channel which are all complex (except the central mode) and so do not necessarily produce real fields at the A-BIC, thus these fields are required to cancel in some way. Fig. \ref{e-field contributions} shows that the weighted basis fields do not have a simple cancellation at the A-BIC. Instead, there are a series of partial cancellations leading to a complete cancellation overall showing that indeed the A-BICs are a result of ``accidental'' destructive interference of leaky modes.

\subsection{Accidental BIC: Varying other parameters}
\label{Sec:Other}

\begin{figure}
	\includegraphics*[clip,width=0.5\textwidth]{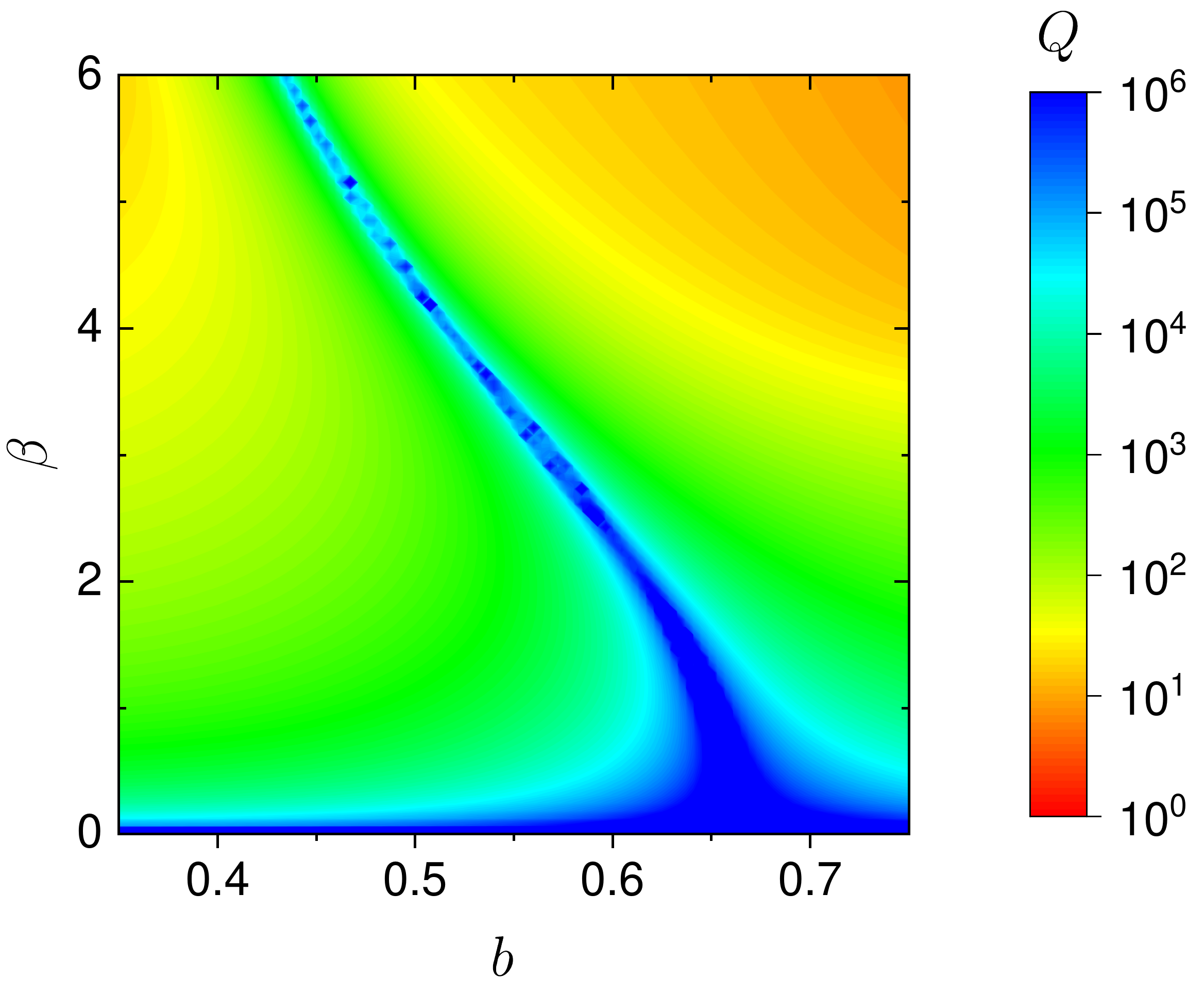}
	\caption{The quality factor $Q$ of the fundamental QGM of the PC slab as a function of the permittivity contrast $\beta$ and the half-width of the perturbation layer, $b$.}
	\label{b-d space}
\end{figure}

We now extend the parameter space while looking for A-BICs to include both the perturbation strength $\beta$ and the perturbation width $b$. By varying both parameters, we need to run the RSE thousands of times, where both the high efficiency and the high accuracy of the RSE become crucial. Focusing on the same fundamental QGM, we plot in \Fig{b-d space} its Q-factor defined as $Q=|\text{Re}\,\omega/(2\text{Im}\,\omega)|$, color-coded on the $\beta$--$b$ plane. Both parameters have natural limits, $b<a$  and $\beta<\varepsilon_s$, which were used in the plot. The latter condition is a requirement that the system stays dielectric, having a positive permittivity, while the former is a fundamental limitation of the RSE that the perturbation must stays within the volume of the basis system. Note, however, that the present version of RSE applied to planar PC structures suffers from the lack of convergence as $b\to a$~\cite{Neale2020}, so that \Fig{b-d space} shows a plot up to $b=0.75a$ only (but the data is reliably calculated up to $b=0.95a$). The Q-factor of the studied mode reaches a numerical value of $Q=10^6$ as it is clear from \Fig{b-d space}, which demonstrates a high accuracy of the RSE calculation of the mode, with the relative error of about $10^{-6}$.

We see from \Fig{b-d space} that an A-BIC is formed at any value of $\beta$, provided that the other parameter is properly tuned, presenting a line in the two-dimensional parameter space. However, no A-BIC is formed for $b>0.7a$ which can be understood as the system does not have thick enough substrate layer (not affected by the periodic modulation) where the destructive interference of leaky modes necessary for A-BIC formation could occur. Clearly, the properties of the A-BIC are also affected by both parameters as can be seen from the strongly inhomogeneous profile of the Q-factor. In fact the range of high Q-values becomes wider as $\beta$ decreases, eventually approaching the limit of the infinite $Q$ for the original guided mode of the homogeneous dielectric slab at $\beta=0$, where $b$ is no longer a variable of the system.

\section{Conclusion}

We have applied the resonant-state expansion (RSE) to planar photonic-crystal structures, in order to find in these systems symmetry protected and accidental bound-states in the continuum (BICs) and to study their properties. We have shown that the eigenvector analysis naturally following from the RSE formalism is a useful tool for identifying BICs, and have provided a rigorous mathematical criterion for differentiating between symmetry-protected and accidental BICs. We have demonstrated how the electro-magnetic field of a resonant state of a photonic-crystal slab can be broken down into homogeneous-slab basis field
components and how those contributions change as the periodic perturbation is modified. In particular, the basis electric fields sum together to create an entirely real field of an accidental BIC, which true to its name is an accidental cancellation of an infinite number of partial leaky waves with no one field fully compensating another. We have furthermore demonstrated that the RSE is an efficient tool for finding the BICs in planar photonic crystal systems due to the unprecedented speed at which it can calculate the modes, thus allowing a multidimensional parameter space to be explored with a high resolution.

\acknowledgments

S.\,N. acknowledges financial support of EPSRC under the DTA scheme. E.\,A.\,M. thanks A.\,F. Sadreev
for discussions.


\begin{thebibliography}{10}
	
	\bibitem{SiegertPR39}
	A.~F.~J. Siegert, Phys. Rev. {\bf 56},  750  (1939).
	
	\bibitem{Weinstein}
	L.~A. Weinstein, {\em Open resonators and open waveguides} (Golem press,
	Boulder, Col., 1969).
	
	\bibitem{WhittakerPRB99}
	D.~M. Whittaker and I.~S. Culshaw, Phys. Rev. B {\bf 60},  2610  (1999).
	
	\bibitem{vonNeumann1929}
	J. von Neumann and E. Wigner, Physikalische Zeitschrift {\bf 30},  467  (1929).
	
	\bibitem{Inoue1982}
	M. Inoue, K. Ohtaka, and S. Yanagawa, Phys. Rev. B {\bf 25},  689  (1982).
	
	\bibitem{Paddon2000}
	P. Paddon and J.~F. Young, Phys. Rev. B {\bf 61},  2090  (2000).
	
	\bibitem{Pacradouni2000}
	V.~Pacradouni, W.~Mandeville, A.~Cowen, P.~Paddon, J.~F.~Young and S.~Johnson , Phys. Rev. B
 {\bf 62},  4204  (2000).
	
	\bibitem{TikhodeevPRB02}
	S.~G. Tikhodeev, A. Yablonskii, E.~A. Muljarov, N.~A Gippius and T. Ishihara, Phys. Rev. B {\bf 66},  045102  (2002).
	
	\bibitem{MarinicaPRL08}
	D.~C. Marinica, A.~G. Borisov, and S.~V. Shabanov, Phys. Rev. Lett. {\bf 100},
	183902  (2008).
	
	\bibitem{BulgakovPRB08}
	E.~N. Bulgakov and A.~F. Sadreev, Phys. Rev. B {\bf 78},  075105  (2008).
	
	\bibitem{Plotnik2011}
	Y. Plotnik, O. Peleg, F. Dreisow, M. Heinrich, S. Nolte, A. Szameit and M. Segev, Phys. Rev. Lett. {\bf 107},  28  (2011).
	
	\bibitem{Hsu2013}
	C.~W. Hsu, B. Zhen, J. Lee, S.~L. Chua, S.~G. Johnson, J.~D. Joannopoulos and M. Solja\v ci\'c, Nature {\bf 499},  188  (2013).
	
	\bibitem{Kodigala2017}
	A. Kodigala, T. Lepetit, Q. Gu, B. Bahari, Y. Fainman and B. Kant\'e, Nature {\bf 541},  196  (2017).
	
	\bibitem{Zhen2013}
	B. Zhen, S. Chua, J. Lee, A.~W. Rodriguez, X. Liang, S.~G. Johnson, J.~D. Joannopoulos, M. Solja\v ci\'c and O. Shapira, Proc. Nat. Acad. Sci. {\bf 110},  13711  (2013).
	
	\bibitem{Yanik2011}
	A.~A. Yanik, A. Cetin, M. Huang, A. Artar, S.~H. Mousavi, A. Khanikaev, J.~H. Connor, G. Shvets and H. Altug, Proc. Nat. Acad. Sci. {\bf 108},  11784  (2011).
	
	\bibitem{Foley2014}
	J.~M. Foley, S.~M. Young, and J.~D. Phillips, Phys. Rev. B {\bf 89},  165111  (2014).
	
	\bibitem{NdangaliJMP10}
	R.~F. Ndangali and S.~V. Shabanov, J. Math. Phys. {\bf 51},
	102901  (2010).
	
	\bibitem{LiSR16}
	L. Li and H. Yin, Sci. Rep. {\bf 6},  26988  (2016).
	
	\bibitem{BulgakovPRL17}
	E.~N. Bulgakov and D.~N. Maksimov, Phys. Rev. Lett. {\bf 118},  267401  (2017).
	
	\bibitem{BulgakovPRA17}
	E.~N. Bulgakov and D.~N. Maksimov, Phys. Rev. A {\bf 96},  063833  (2017).
	
	\bibitem{AzzamPRL18}
	S.~I. Azzam, V.~M. Shalaev, A. Boltasseva, and A.~V. Kildishev, Phys. Rev.
	Lett. {\bf 121},  253901  (2018).
	
	\bibitem{BlanchardPRA14}
	C. Blanchard, P. Viktorovitch, and X. Letartre, Phys. Rev. A {\bf 90},  033824
	(2014).
	
	\bibitem{YangPRL14}
	Y. Yang, C. Peng, Y. Liang, Z. Li and S. Noda, Phys. Rev. Lett. {\bf 113},  037401  (2014).
	
	\bibitem{YoonSR15}
	J.~W. Yoon, S.~H. Song, and R. Magnusson, Sci. Rep. {\bf 5},  18301
	(2015).
	
	\bibitem{BlanchardPRB16}
	C. Blanchard, J.-P. Hugonin, and C. Sauvan, Phys. Rev. B {\bf 94},  155303
	(2016).
	
	\bibitem{NiPRB16}
	L. Ni, Z. Wang, C. Peng, and Z. Li, Phys. Rev. B {\bf 94},  245148  (2016).
	
	\bibitem{Sadrieva2017}
	Z.~F. Sadrieva , I. Sinev, K.~L. Koshelev, A. Samusev, I.~V. Iorsh, O. Takayama, R. Malureanu, A.~A. Bogdanov and A.V. Lavrinenko , ACS Photonics {\bf 4},  723  (2017).
	
	\bibitem{YuanOL17}
	L. Yuan and Y.~Y. Lu, Opt. Lett. {\bf 42},  4490  (2017).
	
	\bibitem{BulgakovPRA18}
	E.~N. Bulgakov and D.~N. Maksimov, Phys. Rev. A {\bf 98},  053840  (2018).
	
	\bibitem{BykovPRA19}
	D.~A. Bykov, E.~A. Bezus, and L.~L. Doskolovich, Phys. Rev. A {\bf 99},  063805
	(2019).
	
	\bibitem{OvcharenkoPRB20}
	A.~I. Ovcharenko, C. Blanchard, J.-P. Hugonin, and C. Sauvan, Phys. Rev. B {\bf
		101},  155303  (2020).
	
	\bibitem{BulgakovPRA14}
	E.~N. Bulgakov and A.~F. Sadreev, Phys. Rev. A {\bf 90},  053801  (2014).
	
	\bibitem{Hsu2016}
	C.~W. Hsu, B. Zhen, A.~D. Stone, J.~D. Joannopoulos and M. Solja\v ci\'c, Nature Reviews Materials {\bf 1}, 16048  (2016).
	
	\bibitem{GaoSR16}
	X. Gao, C.~W. Hsu, B. Zhen, X. Lin, J.~D. Joannopoulos, M. Solja\v ci\'c and H. Chen, Sci. Rep. {\bf 6},  31908  (2016).
	
	\bibitem{Pilipchuk2017}
	A.~S. Pilipchuk and A.~F. Sadreev, Phys. Lett. A {\bf 381},  720  (2017).
	
	\bibitem{MuljarovEPL10}
	E.~A. Muljarov, W. Langbein, and R. Zimmermann, Europhys. Lett. {\bf 92},  50010
	(2010).
	
	\bibitem{DoostPRA14}
	M.~B. Doost, W. Langbein, and E.~A. Muljarov, Phys. Rev. A {\bf 90},  013834
	(2014).
	
	\bibitem{LobanovPRA17}
	S.~V. Lobanov, G. Zoriniants, W. Langbein, and E.~A. Muljarov, Phys. Rev. A
	{\bf 95},  053848  (2017).
	
	\bibitem{DoostPRA12}
	M.~B. Doost, W. Langbein, and E.~A. Muljarov, Phys. Rev. A {\bf 85},  023835
	(2012).
	
	\bibitem{DoostPRA13}
	M.~B. Doost, W. Langbein, and E.~A. Muljarov, Phys. Rev. A {\bf 87},  043827
	(2013).
	
	\bibitem{LobanovPRA19}
	S.~V. Lobanov, W. Langbein, and E.~A. Muljarov, Phys. Rev. A {\bf 100},  063811
	(2019).
	
	\bibitem{ArmitagePRA14}
	L.~J. Armitage, M.~B. Doost, W. Langbein, and E.~A. Muljarov, Phys. Rev. A {\bf
		89},  053832  (2014).
	
	\bibitem{ArmitagePRA18}
	L.~J. Armitage, M.~B. Doost, W. Langbein, and E.~A. Muljarov, Phys. Rev. A {\bf
		97},  049901  (2018).
	
	\bibitem{MuljarovOL18}
	E.~A. Muljarov and T. Weiss, Opt. Lett. {\bf 43},  1978  (2018).
	
	\bibitem{WeissPRL16}
	T. Weiss, M. Mesch, M. Sch\"aferling, H. Giessen, W. Langbein and E.~A. Muljarov, Phys. Rev. Lett. {\bf 116},  237401  (2016).
	
	\bibitem{Neale2020}
	S. Neale and E.~A. Muljarov, Phys. Rev. B {\bf 101},  155128  (2020).
	
	\bibitem{WeissPRB17}
	T. Weiss, M. Sch\"aferling and H. Giessen, Phys. Rev. B {\bf 96},  045129  (2017).
	
	\bibitem{LobanovPRA18}
	S.~V. Lobanov, W. Langbein, and E.~A. Muljarov, Phys. Rev. A {\bf 98},  033820
	(2018).
	
	\bibitem{GrasOL19}
	A. Gras, W. Yan, and P. Lalanne, Opt. Lett. {\bf 44},  3494  (2019).
	
	\bibitem{MuljarovPRB16Purcell}
	E.~A. Muljarov and W. Langbein, Phys. Rev. B {\bf 94},  235438  (2016).
	
\end{thebibliography}

\end{document}